\documentclass[letterpaper,twoside]{article}
\pdfoutput=1 

\usepackage[top=2.5cm, bottom=2.5cm, left=3cm, right=3cm]{geometry}
\usepackage{graphicx}
\DeclareGraphicsExtensions{.pdf,.png,.jpg,.mps,.eps,.ps}
\graphicspath{{../figures/}{.}}
\usepackage[numbers,sort&compress,super,comma]{natbib}
\usepackage[table]{xcolor}
\usepackage{booktabs}       
\usepackage{multirow}
\usepackage{enumitem}
\usepackage{hyperref}
\hypersetup{
    colorlinks = true,
    citecolor = .,      
    linkcolor = .
    }
\usepackage{url}

\newcommand{\lcglyph}{\protect\inlinegraphics{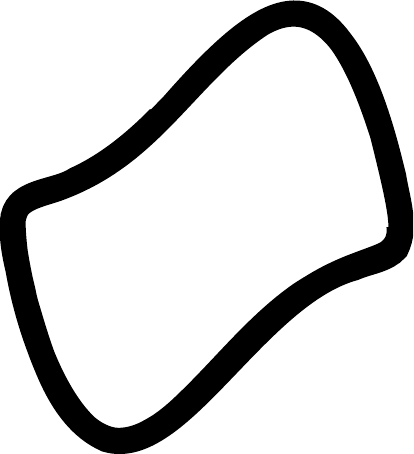}}

\usepackage{microtype} 
\usepackage[T1]{fontenc}
\usepackage[default]{lato} 
\usepackage{fancyhdr}
\usepackage{titling}

\pretitle{\begin{center}\LARGE}
\posttitle{\end{center}}
\preauthor{\begin{center}
\large \lineskip 0.5em}
\postauthor{\par\end{center}
\begin{center}
\textit{
$^1$Champalimaud Centre for the Unknown, Champalimaud Foundation, Lisbon, Portugal
\\
$^2$Department of Neurobiology and Behavior, Stony Brook University, NY, USA
}
\\
$^\ast$corresponding author, \url{memming.park@research.fchampalimaud.org}
\end{center}}

\title{Persistent learning signals and working memory\\without continuous attractors}
\author{I. Memming Park$^{1,2,\ast}$, \'Abel S\'agodi$^{1}$, and Piotr A.~Sok\'o\l$^{1,2}$}
\date{}

\usepackage[fleqn]{amsmath} 
\usepackage{amssymb}
\usepackage{mathtools} 
\usepackage{bm} 

\usepackage{letltxmacro}
\LetLtxMacro{\originaleqref}{\eqref}
\renewcommand{\eqref}{Eq.~\originaleqref}
\definecolor{c:adjointidx}{rgb}{0.15,0.55,0.7}
\definecolor{c:time}{rgb}{0.50,0.12,0.7}

\usepackage{forloop}
\newcommand{\defvec}[1]{\expandafter\newcommand\csname v#1\endcsname{{\mathbf{#1}}}}
\newcounter{ct}
\forLoop{1}{26}{ct}{
    \edef\letter{\alph{ct}}
    \expandafter\defvec\letter
}

\forLoop{1}{26}{ct}{
    \edef\letter{\Alph{ct}}
    \expandafter\defvec\letter
}

\DeclarePairedDelimiter{\abs}{\lvert}{\rvert}
\DeclarePairedDelimiter{\norm}{\lVert}{\rVert}

\newcommand{\dm}[1]{\ensuremath{\mathrm{d}{#1}}} 
\newcommand{\RN}[2]{\frac{\dm{#1}}{\dm{#2}}} 
\newcommand{\PD}[2]{\frac{\partial #1}{\partial #2}} 

\newcommand{\field}[1]{\ensuremath{\mathbb{#1}}}
\newcommand{\reals}{\field{R}}

\newcommand{\trp}{{^\top}}

\newcommand{\DF}{\nabla_{\vx}\vf}
\newcommand{\vDelta}{\mathbf{\Delta}}
\newcommand{\vPsi}{\mathbf{\Psi}}
\newcommand{\vphi}{\bm{\phi}}

\newlength\myheight
\newlength\mydepth
\settototalheight\myheight{Xygp}
\newcommand*\inlinegraphics[1]{%
  \settototalheight\myheight{Xygp}%
  \settodepth\mydepth{Xygp}%
  \raisebox{-\mydepth}{\includegraphics[height=\myheight]{#1}}%
}

\begin{document}
\thispagestyle{fancy}
\maketitle
\thispagestyle{fancy}

\begin{abstract}
Neural dynamical systems with stable attractor structures, such as point attractors and continuous attractors, are hypothesized to underlie meaningful temporal behavior that requires short-term/working memory.
However, working memory may not support useful learning signals necessary to adapt to changes in the temporal structure of the environment.
We show that in addition to the continuous attractors that are widely implicated, periodic and quasi-periodic attractors can also support learning arbitrarily long temporal relationships.
Unlike the continuous attractors that suffer from the fine-tuning problem, the less explored quasi-periodic attractors are uniquely qualified for learning to produce temporally structured behavior.
Our theory has broad implications for the design of artificial learning systems and makes predictions about observable signatures of biological neural dynamics that can support temporal dependence learning and working memory.
Based on our theory, we developed a new initialization scheme for artificial recurrent neural networks that outperforms standard methods for tasks that require learning temporal dynamics.
Moreover, we propose a robust recurrent memory mechanism for integrating and maintaining head direction without a ring attractor.
\end{abstract}

\section{Introduction}
Exploiting the temporal structure of the world is essential for an agent to produce meaningful behaviors.
Learning long-range temporal dependencies, where there is a temporal separation between the production of the desirable behavior and the presentation of the relevant information, is a major challenge for both biological and artificial neural systems.
Agents are often required to digest the information over time and are rewarded for producing timed behavior in a controlled research environment.
For example, in perceptual decision-making tasks, noisy sensory stimuli over time are integrated to generate the reported decisions~\cite{Gold2001}.
Many tasks in neuroscience, cognitive sciences, and machine learning involve timing, working memory, and temporal integration, such as delay eyeblink conditioning, delayed discrimination, and random delay copy memory tasks~\cite{Arjovsky2016,Giovannucci2017,Premereur2011}.
Moreover, the timing structure of these tasks demands temporal flexibility and generalization; that is, there is no expected limit in the relevant temporal extent---for example, the subject is expected to maintain its performance when the working memory duration of a task is longer than that encountered during training.

Here, we consider learning temporal structures through incremental changes to the parameters (e.g., synaptic weights) using the gradient signal that informs the learning system how to change towards a particular direction that improves performance.
Since the days of Rosenblatt's Perceptron (1957) and Widrow and Hoff's least mean squares algorithm (1960), learning guided by the gradients of the error had an enormous impact on the connectionist theories of how biological neural system learn~\cite{Widrow1990}.
It led to the widespread use of artificial neural networks, and eventually to the deep learning revolution and modern artificial intelligence.
Nowadays, gradient-based learning is established as the mainstream numerical technique for learning and optimizing complex artificial systems, while putative neurobiological implementations that go beyond a single feedforward layer have been long sought after~\cite{Richards2019,Bellec2020,Lillicrap2020,Richards2023}.

In general, gradient-based learning applied to solving tasks with temporal structures requires a memory mechanism that maintains and propagates the learning signal across temporal gaps.
However, the theoretical mechanism that supports learning signals over long periods of time remains a mystery because the seemingly simple persistent activity solutions are at odds with gradient-based learning as we shall explore in depth.

Research in working memory mechanisms provides a useful framework for bridging temporal gaps.
Analyzing working memory as a neural dynamical system, three types of dynamical structures have been generally considered in the literature: fading memory, multiple point attractors, and continuous attractors.
Fading memory system which includes functionally feedforward and reservoir architectures~\cite{Waibel1989,deVries1992,Maass2002,Goldman2009,Jaeger2007,Vaswani2017} store information in the transient population activity that eventually returns to a globally stable state.
This fading memory property imposes a fundamental limit on their maximum temporal extent and is not suitable, at least theoretically, for an arbitrarily long time scale.
Multiple isolated point attractors can persistently maintain neural state at any one of the point attractors, hence can store discrete information indefinitely~\cite{Hopfield1984,Bengio1994,Wang2002,Brinkman2021a}.
Continuous attractor stores information on a continuous manifold such that the neural state can be maintained to represent an arbitrary continuous value for an arbitrary duration~\cite{Seung1996,Machens2005,Khona2022}.
While both the point and continuous attractor mechanisms support working memory---necessary for causal learning over an unrestricted interval, point attractors produce unsuitable learning signals such that the change from one time point to another tends to decay exponentially quickly as the interval increases.

In this paper, we present a general mathematical theory of gradient learning signal propagation through time in recurrent networks to analyze their relation to neural dynamical structures.
This work not only challenges previous models of biological working memory and temporal dependence learning but also informs the design of machine learning algorithms.
By focusing on the long time scale by taking the asymptotic time limit, we can greatly simplify the analysis since the results do not depend on the fine details of dynamics but rather only on the global topological structures of the neural dynamics.
We argue that the persistence of asymptotic learning signals---persistence in the sense that they exhibit neither vanishing nor exploding behavior---is a necessary condition to practically learn using gradients.
At the same time, we demand that the neural dynamics that support learning to be robust to small perturbations in the parameters.
We show that current models of memory fail to satisfy both criteria as they either do not support unbounded temporal learning or are not robust to parameter perturbations.
For example, it is well known that the continuous attractor model suffers from the fine-tuning problem: Small changes in the system parameters almost always destroys the continuous attractor feature~\cite{Seung1996}.
We propose an alternative recurrent mechanism, \mbox{(quasi-)}periodic attractors, with learning signals that are both robust and asymptotically persistent, thus satisfying both requirements for unbounded temporal learning.
The proposed mechanism encodes information in oscillations, which leads us to hypothesize about the oscillatory activity that occurs at multiple temporal and spatial scales in the brain.
To demonstrate its practical benefits, we devised an initialization scheme for artificial recurrent neural networks (RNN) that enables learning challenging tasks.
We conjectured that the (quasi-)periodic attractors are the only dynamical structure that are both robust and asymptotically appropriate for temporal learning.
Finally, we propose a novel continuous working memory mechanism that behaves like the continuous ring attractor and can produce a persistent activity bump without a traditional continuous attractor.

\section{Results}
\subsection{Physical limits of gradient signal representation}\label{sec:physical_limits}
Let us briefly define gradient based learning.
Let $w_k$ be the $k$-th adjustable parameter (e.g. synaptic weight) in the chain of information processing from the source (e.g. sensory stimuli) to output (e.g. behavioral report).
In the context of learning a task, the task design is such that certain behaviors are considered desirable and any deviation is considered to be an error.
Let $L$ denote the ``loss'', a numerical quantification of the magnitude of the error for generating a certain output corresponding to an input.
The gradient signal is the partial derivative of the loss with respect to each parameter:
$$
\frac{\partial L}{\partial w_k} = \lim_{\Delta w \to 0} \frac{L(w_k + \Delta w) - L(w_k)}{\Delta w}.
$$
As the definition indicates, positive $\frac{\partial L}{\partial w_k}$ implies an infinitesimal decrease in the parameter value $w_k$ decreases the amount of error measured by the loss $L$.
Given the gradient of all adjustable parameters, we can adjust all of them proportional to the negative of the gradients to decrease the loss $L$ in the steepest direction.
For example, we can use \emph{gradient flow} that defines the continuous dynamics of the weights, or the discrete step-like analogue used in machine learning:
\begin{align}
    \label{eq:gradient-flow}
    \frac{\mathrm{d}w_k}{\mathrm{d}t} &= -\frac{1}{\tau} \frac{\partial L}{\partial w_k}
&\qquad \text{(gradient flow)}
    \\
    \label{eq:gradient-descent}
    w_k &\leftarrow w_k -\frac{\Delta t}{\tau} \frac{\partial L}{\partial w_k}
&\qquad \text{(gradient descent)}
\end{align}
simultaneously for all parameters.
The positive constant $\tau$ is the time constant for learning, and its reciprocal $\frac{1}{\tau}$ is the learning rate (per unit time).
While this particular method of learning uses the principle of gradient descent, the theory we develop offers insights that are not limited to a particular implementation of gradient descent~\cite{Minsky1988,Richards2023} nor does it depend on the nature of the learning signal, i.e., it need not derive from a supervised learning problem.
More broadly, our analysis focuses on the quality of the learning signal itself---%
locally, gradient descent offers an optimal reduction in the loss and as such provides a useful baseline to compare other models of learning.

Mathematically, as long as every information processing component of the system and the loss function are differentiable, calculus (chain rule) provides a foundation for the necessary gradients (see Discussion for generalization and bioplausibility, Sections \ref{sec:gengrad} and \ref{sec:biorelevance}).
Note that there is no limit to the magnitude of the gradients in theory---in magnitude, gradients can be arbitrarily small (close to zero) or arbitrarily large.
However, in practice, since the gradients have to be represented as physical states efficiently by the learning system, the gradients that are too small or too large in magnitude pose major challenges in learning.
If the gradients are directly represented as (bio)physical quantities (e.g. ion concentrations or membrane potential), small gradient values can go below the noise floor and get lost, while large gradient values cannot be represented due to the finite nature of the physical system.
On the other hand, abstract number systems used in digital computers can only efficiently handle a fixed range of numbers with the usual floating point representation.
Thus, in both cases, there is an effective range where gradient descent can be realized.
Moreover, gradients that are too large can be especially harmful for learning, as the learning trajectory and system behavior can become unpredictable in practice.

Unfortunately, gradients that are too small or too large in magnitude are commonly encountered, especially in recurrent networks and deep neural networks, a phenomenon dubbed the Exploding and Vanishing Gradient Problem (EVGP).
EVGP causes training very deep learning architectures inefficient and renders impractical.
Therefore, EVGP has been a focus of a large body of research in machine learning with many practical interventions~\cite{Hochreiter1997,Hinton2006,Ozturk2007,Srivastava2015,He2016,Ioffe2015,Mhammedi2017,Arjovsky2017,Rusch2021} and theoretical considerations~\cite{Doya1993,Bengio1994,Hochreiter2001,Glorot2010,Pascanu2013,Saxe2014,Hanin2018,Pennington2017a,Sokol2018a}.
Since having a healthy learning signal is a necessary condition for gradient-based learning for both biological and artificial systems, our goal is to theoretically analyze and avoid EVGP.
In the next section, we develop a general mathematical theory of gradient signal propagation in recurrent networks.

\subsection{Sensitivity, adjoint, and gradient propagation through time}\label{sec:sagptt}

\begin{figure}[tbhp]
    \centering
    \includegraphics{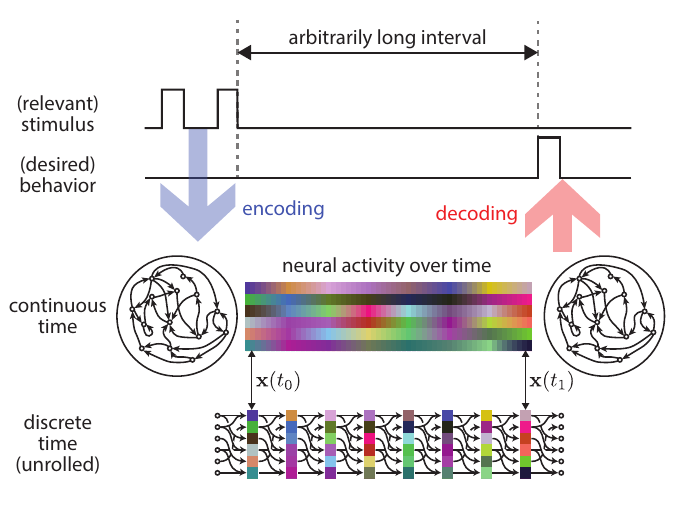}
    \caption{
	\textbf{Recurrent activity in the network must support all necessary information through time.}
	(Top) Relevant stimulus input is encoded as neural activity pattern in the past.
	Desired behavioral output is decoded from neural activity pattern in the future.
	The interval between the two can be arbitrarily large, making it difficult to learn.
	(Middle) Continuous time evolution of finite dimensional internal neural representation.
	(Bottom) Discrete-time approximation of the continuous time process for artificial recurrent neural networks and computational modeling.
    }
    \label{fig:rnn:diagram}
\end{figure}

What is the \emph{fundamental source} of vanishing or exploding gradient signals?
In recurrent networks, we can easily gain intuition if we view them as a dynamical system, or an ordinary differential equation that describes how the neural state $\vx(t)$ evolves over time~\cite{Cowan1972,Baldi1995,Beer1995}:
\begin{align}\label{eq:ode}
\frac{\mathrm{d} \vx}{\mathrm{d}t} = \vf(\vx(t), \vu(t); \vw),
\end{align}
where $\vx(t) \in \reals^n$ denotes the vector of internal neural activations, $\vu(t) \in \reals^{d_u}$ denotes the external input at time $t$, and $\vf(\cdot, \cdot\, ; \vw) \in \reals^n \times \reals^{d_u} \to \reals^n$ is a function that defines the vector field~\cite{Chicone2006,Izhikevich2007}.
All learnable parameters are denoted by the vector $\vw$.
Discrete-time recurrent neural networks can be seen as a numerical integration of a corresponding continuous time system (Fig.~\ref{fig:rnn:diagram}).
The main advantage of the ordinary differential equation of the form \eqref{eq:ode} is the access to the powerful tools of continuous dynamical systems theory.
Note that the $n$-dimensional network activations $\vx(t)$ is the sole carrier of information across time.
If information about the input in the past is needed, it has to have been stored in an internal memory, a memory accessible through decoding the instantaneous pattern of activation $\vx(t)$ in the future (Fig.~\ref{fig:rnn:diagram}).

\begin{figure}[tbhp]
    \centering
    \includegraphics[width=\textwidth]{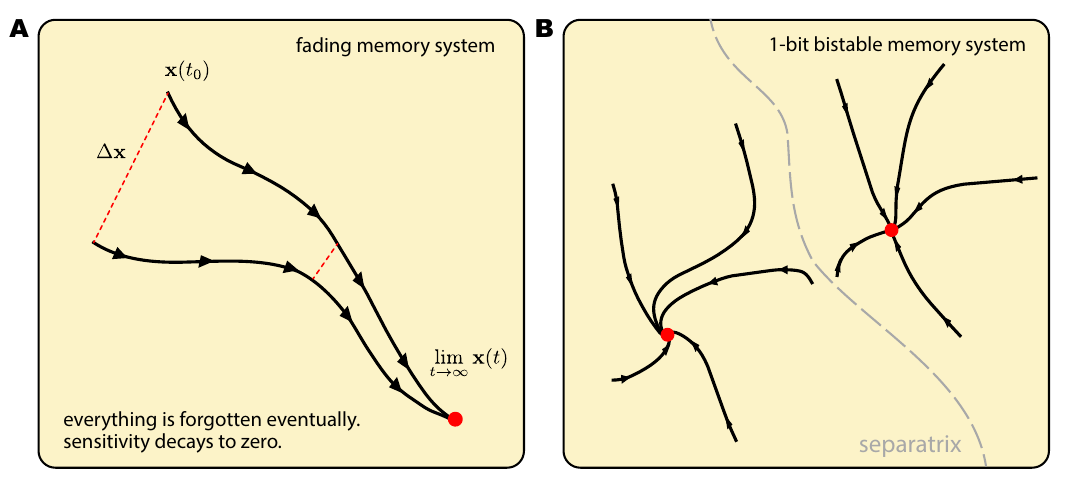}
    \caption{
	\textbf{(A)}
	\textbf{Fading memory dynamics forgets small and large perturbations asymptotically.}
	Both $\vx(t_0)$ and its perturbation $\vx(t_0) + \Delta \vx$ evolves over time and eventually gets close to the same global stable fixed point (red).
	Any memory encoded in the neural activation vector and the corresponding adjoint are forgotten.
	Although the neural state vector is 2-dimensional in this illustration, the principle is true for any fading memory system.
	\textbf{(B)}
	\textbf{One bit memory but vanishing gradients.}
	There are two isolated stable point attractors with their corresponding basins of attraction.
	Similar to (A), the asymptotic sensitivity vanishes, however, the identity of the basin of attraction remains.
	This example illustrates the decoupling of persistent memory and persistent learning signal propagation over a long time.
    }
    \label{fig:sensitivity}
    \label{fig:dyn:globalfp}
    \label{fig:dyn:twobasins}
\end{figure}

Let us consider a recurrent network with a single global resting state.
Any fading memory system eventually forgets the past, since any difference $\Delta$ between any two neural activation patterns eventually tends zero as they both evolve to the same resting state (Fig.~\ref{fig:dyn:globalfp}A).
At the same time, the ability of the dynamical system to carry \emph{small perturbations} of the past state $\vx(t_0)$ to produce an impact in the future state $\vx(t_1)$ acts as a bottleneck for the gradient signals.
Consider the simple case where the loss is only dependent on the neural activations at some time $t_1$.
Given a neural trajectory over time $\vx(t)$ from $t_0$ to $t_1$, we can decompose the gradient of the loss as the integral of all possible paths of information flow among the $n$ neurons from any time $t$ in the past:
\begin{align}\label{eq:grad:loss}
    \frac{\partial L}{\partial w_k} &=
	\sum_{i,j,l}
	\int_{t_0}^{t_1}
	\frac{\partial L}{\partial x_i(t_1)}
	\underbrace{\frac{\partial x_i(t_1)}{\partial x_j(t)}}_{\substack{\text{adjoint}}}
	\frac{\partial x_j(t)}{\partial f_l(\vx(t), \vu(t); \vw)} \frac{\partial f_l(\vx(t), \vu(t); \vw)}{\partial w_k}
	\mathrm{d}t,
\end{align}
where $x_i(t)$ denotes the $i$-th component of the $n$-dimensional vector $\vx(t)$, and likewise for $f_l(\cdot)$ for $\vf(\cdot)$.
This is a continuous analog of the \emph{backpropagation-through-time} algorithm~\cite{Sanz-Serna2016,Toomarian1991,Huh2018,Gholami2019}.
Importantly, the only term that stretches along with the temporal interval between $t_0$ and $t_1$ is the term annotated as the \emph{adjoint}, also known as the backward sensitivity~\cite{Pontryagin1964,Brockett1970,Ermentrout2010}.
The time evolution of the adjoint is central to understanding the EVGP.

The adjoint is tightly related to the (forward) sensitivity, as can be seen from the symmetry of their definitions:
\begin{align}
    \psi_{i,j}({\color{c:time}t}) &= \frac{\partial x_j(t_1)}{\partial x_i({\color{c:time}t})} = \lim_{\Delta x_i \to 0} \frac{x_j(t_1, \vx({\color{c:time}t})+\Delta x_i) - x_j(t_1, \vx({\color{c:time}t}))}{\Delta x_i}
    \qquad & \text{adjoint}
    \label{eq:adjoint}
    \\
    \delta_{i,j}({\color{c:time}t}) &= \frac{\partial x_i({\color{c:time}t})}{\partial x_j(t_0)} = \lim_{\Delta x_j \to 0} \frac{x_i({\color{c:time}t}, \vx(t_0)+\Delta x_j) - x_i({\color{c:time}t}, \vx(t_0))}{\Delta x_j}
    \qquad & \text{(forward) sensitivity}
    \label{eq:sensitivity}
\end{align}
where we explicitly expressed the dependence of neural trajectory $\vx(t)$ with respect to its initial condition at $t_0$ as $\vx(t, \vx(t_0))$.
The sensitivity quantifies how much influence an infinitesimal perturbation at time $t_0$ has on the future times $t$.
Note that if the sensitivity or the adjoint is zero, it indicates a disconnection in the chain of derivatives, and thus an absence of temporal learning signal between the neuron $i$ and $j$.
The adjoint and sensitivity evolves over time in a complementary fashion such that their inner product stays constant over time~\cite{Pontryagin1964}.
\begin{align}\label{eq:hamiltonian}
    \sum_{k} \psi_{k,i}(t) \delta_{k,j}(t)
    =
	\sum_{k}
	\frac{\partial x_i(t_1)}{\partial x_k(t)}
	\frac{\partial x_k(t)}{\partial x_j(t_0)}
	=
	\frac{\partial x_i(t_1)}{\partial x_j(t_0)}
\end{align}
for all $i,j$.
Or equivalently expressed in a matrix form,
\begin{align}
    \underbrace{\vPsi^\top_{t_1}(t)}_{\substack{\text{adjoint}}}
    \underbrace{\vDelta_{t_0}(t)}_{\substack{\text{sensitivity}}}
    &=
    \underbrace{\vJ_{t_0}^{t_1}}_{\substack{\text{time-time Jacobian}}}
\end{align}
where the $(n \times n)$ matrices correspond to
$\left[\mathbf{\Psi}_{t_1}\right]_{i,j} = \psi_{i,j}(t)$,
$\left[\mathbf{\Delta}_{t_0}\right]_{i,j} = \delta_{i,j}(t)$,
$\left[\vJ_{t_0}^{t_1}\right]_{i,j} = \frac{\partial x_i(t_1)}{\partial x_j(t_0)}$,
and $(\cdot)^\top$ denotes matrix transpose operation.
Since by definition $\mathbf{\Psi}^\top_{t}(s) = \mathbf{\Delta}_s(t)$,
given an additive perturbation of the initial state $\vx(t_0)$ by a vector $\vv$,
the asymptotic time evolution of the norms share the same fate:
\begin{align}\label{eq:asymptotic:dual}
    \lim_{t_1 \to \infty}
    \norm[\Big]{\vDelta_{t_0}(t_1) \vv}
    =
    \lim_{t_1 \to \infty}
    \norm[\Big]{\vPsi^\top_{t_1}(t_0) \vv}
\end{align}
as the time interval $(t_1 - t_0)$ goes to infinity.
Hence, if the sensitivity in the direction of $\vv$ at time $t_0$ decays to the origin, the adjoint also decays to the origin~\cite{Brockett1970}.
Similarly, if the sensitivity diverges to infinity, so does the adjoint~\cite{Brockett1970}.
The EVGP phenomenon is exacerbated for longer durations or deeper networks, precisely because of the converging or diverging behavior of the adjoint in~\eqref{eq:grad:loss}.

Even though the above analysis is for a particular trajectory of neural activations, the asymptotic behavior of the sensitivity (and therefore the adjoint) is strongly determined by the topological structure of the dynamics of the system, if all trajectories share the same fate.
For instance,
for fading memory systems in general (depicted in Fig.~\ref{fig:sensitivity}A), as the size of the perturbation at the initial state approaches infinitesimally small, the difference at later time corresponds to the (forward) sensitivity (\eqref{eq:sensitivity}), due to the global point attractor topology.
Coinciding with the memory content that fades away, the sensitivity tends to zero in the limit of long duration, exhibiting \emph{asymptotically vanishing gradients}.
Thus, arbitrarily long-temporal relations cannot be learned by relying on temporally propagated gradients, a difficulty well-known to a large class of theoretical and practical systems.
Notably, reservoir computing frameworks such as echo state networks~\cite{Jaeger2007} and liquid state machines~\cite{Maass2002}, and any stable, linear dynamical system models of neural computation fall in this class. 

\subsection{Robust binary memory with vanishing gradient signal}\label{sec:decouplingmemory}
Consider a system with two basins of attraction each with a point attractor\footnote{also known as stable equilibrium or stable fixed point} (Fig.~\ref{fig:dyn:twobasins}B).
In other words, any initial neural activation vector in one of the basins decays to the corresponding equilibrium.
In this bistable system, similar to the globally stable point attractor system (Fig.~\ref{fig:dyn:globalfp}A), small perturbations are almost always forgotten, and hence the sensitivity also decays asymptotically (i.e., causes vanishing gradient).
But globally, the bistability can be used to store 1-bit of information by the identity of the basin of attraction---for example, which of the two alternative input classes were presented, or which of the two alternative decisions were made~\cite{Wang2002}---indefinitely.
Therefore, it can function as a perfect working memory system, but not ideal for learning arbitrarily long temporal relationships.
This dissociation generalizes to any number of stable equilibria implementing a robust discrete memory system but with vanishing gradients~\cite{Bengio1994}.

Long-term associative memory structures that rely on attractor dynamics also suffer from the same consequence.
For example, Hopfield networks stores each memory item with a corresponding stable attractor such that incomplete or noisy representation of memories within the basin of attraction can recall the identity of the attractor~\cite{Hopfield1984,Amit1989}.
Gradient-based learning signals over time in trained Hopfield networks necessarily vanishes asymptotically, making updating the network in realistic scenarios where the stimulus presentation and the reward signals are temporally separated difficult.
Therefore, to reconfigure recurrent systems that learned to utilize stable fixed points (e.g., to learn a new task), the learning procedure would benefit from incorporating mechanisms that do not rely solely on the gradients of the loss function for performance, such as noise injection, Hebbian plasticity, or neuromodulation (homeostasis) of the depth of attractors (transient or permanent forgetting).

\subsection{Lyapunov spectrum characterizes the asymptotic learning signals}\label{sec:lyapunov}
For a general multi-dimensional dynamical system other than point attractors, depending on the direction of perturbation, the evolution of the perturbed state (i.e., trajectory of the neural activation vector) may have different consequences.
Therefore, we need a multi-dimensional notion of the asymptotic behavior of sensitivity.
The sensitivity and the adjoint dynamics can be expressed through the linearization of the ordinary differential equation~\eqref{eq:ode} using the Jacobian matrix of the dynamics $\vf(\vx, \vu; \vw)$ with respect to the neural state vector:
\begin{align}\label{eq:lyapunov:jacobian}
    \DF(\vx, \vu) &\coloneqq
	\left.
	\begin{bmatrix}
	    \PD{f_1}{x_1} & \PD{f_1}{x_2} & \cdots & \PD{f_1}{x_n} \\
	    \vdots & \vdots & \ddots & \vdots \\
	    \PD{f_n}{x_1} & \PD{f_n}{x_2} & \cdots & \PD{f_n}{x_n} \\
	\end{bmatrix}
	\,
	\right\vert_{(\vx, \vu, \vw)}
	\in
	\reals^{n \times n}
    \qquad \text{(Jacobian of the flow)}
\end{align}
Given a neural trajectory $\vphi(t) \coloneqq \vx(t, \vx(t_0), \vu)$, we can linearize the nonlinear dynamics given by \eqref{eq:ode} along the trajectory which describes the dynamics of the corresponding sensitivity and adjoint matrices~\cite{Brockett1970}:
\begin{align}\label{eq:variational}
    \RN{\vDelta_{t_0}(t)}{t} = \DF(\vphi(t), \vu(t))\vDelta_{t_0}(t),
\qquad
    \RN{\vPsi_{t_1}(t)}{t} = -\DF(\vphi(t), \vu(t))\trp\vPsi_{t_1}(t)
\qquad
\end{align}

\noindent
To illustrate its implications, consider a linear system:
\begin{align}\label{eq:lyapunov:linear:2D}
    \RN{\vx}{t}
	&= \vf(\vx)
	= \vA \vx
\end{align}
for which the Jacobian $\DF(\vphi(t), \vu(t)) = \vA$ does not depend on the trajectory, input, nor time.
Thus the sensitivity dynamics is simply,
\begin{align}
    \RN{\vDelta_{t_0}(t)}{t} = \vA \vDelta_{t_0}(t).
\end{align}
Therefore, the eigenvalues of $\vA$ determine the temporal evolution of the learning signal no matter what the current neural state is. 
The number of positive, negative, and zero eigenvalues determine the dimension of the subspaces corresponding to exploding, vanishing, and sub-exponential behavior.
Therefore, having zero eigenvalues in $\vA$ is a necessary condition for avoiding the asymptotic EVGP on the corresponding subspace.
In the simplest case of normal matrix $\vA$ without algebraic multiplicities, for a perturbation in the direction of the $i$-th eigenvector $\ve_i$ corresponding to the eigenvalue $\lambda_i$, the norm of the sensitivity evolves as,
\begin{align}
    \norm[\Big]{
	\vDelta_{t_0}(t) \ve_i
    }
    \propto
    \norm[\Big]{
	e^{\vA t}\ve_i
    }
    =
    e^{\lambda_i t}.
\end{align}
which is constant when $\lambda_i = 0$.
A similar argument on learning signal holds for a general matrix $\vA$.
Unfortunately, extending the notion of exponential rates of convergence and divergence to nonlinear dynamical systems is trickier than simply inspecting the eigenvalues of the time-varying Jacobian~\cite{Josic2008}, and thus, we need the notion of \emph{Lyapunov exponents} which coincides with the eigenvalues for \eqref{eq:lyapunov:linear:2D}.
The Lyapunov exponent $\Lambda$ is a real number that measures exactly the exponential rate of the (asymptotic) sensitivity dynamics~\cite{Chicone2006}:
\begin{align}\label{eq:lyapunov:exponent}
\Lambda(\vx(t_0), \vv) &= \limsup_{t \to \infty}
    \frac{1}{t} \log
	\left\Vert
		\frac{\partial \vx(t)}{\partial \vx(t_0)}\vv
	\right\Vert
\end{align}
where the unit vector $\vv$ denotes the initial perturbation direction.
If $\Lambda(\vx(t_0),\vv)$ is negative, then the sensitivity (and the adjoint) asymptotically vanishes at the corresponding rate, and if it is positive, the sensitivity asymptotically explodes~\cite{Vogt2022}.
Chaotic dynamics have at least one positive Lyapunov exponent (but the converse is not true), hence prone to the exploding gradient problem~\cite{Mikhaeil2022}.
These properties of learning signals have been studied in the context of random networks~\cite{Chen2018-mf}, and chaos~\cite{Engelken2020,Mikhaeil2022}.

\begin{figure}[tbhp]
    \centering
    \includegraphics[width=\textwidth]{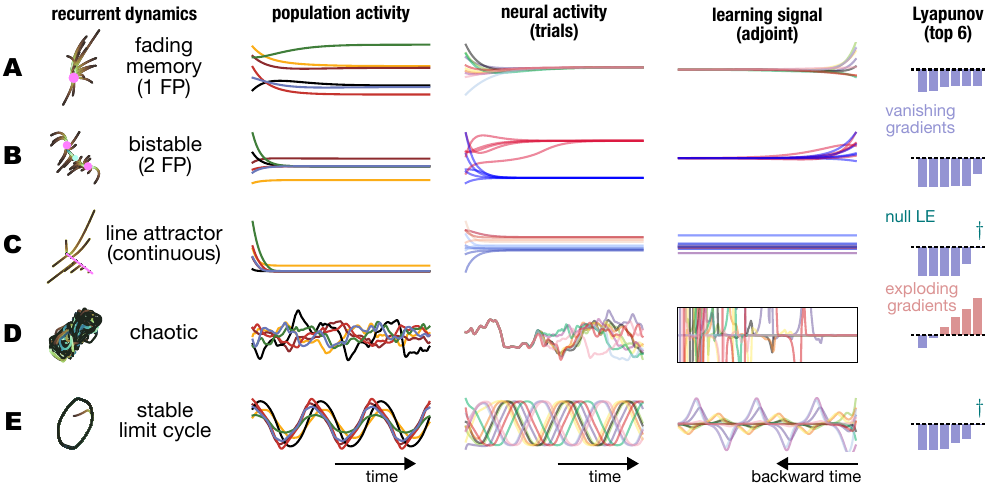}
    \caption{
	\textbf{Neural activity, adjoint, and the Lyapunov spectrum corresponding to 5 stereotypical dynamical systems.}
	\textbf{(A)}
	Single global stable fixed point dynamics representing fading memory systems (Sec.~\ref{sec:sagptt}).
	Example trajectories visualized with principal components analysis shows that regardless of initial condition they converge to the pink fixed point (FP).
	The population activity for a single initialization shows the convergence of each neurons activity overt time.
	The neural activity of a single neuron for different initializations (color) shows spontaneous decay over time.
	The adjoint signal that captures the decaying gradient over backward time corresponding to an error signal (desired perturbation) at the final time.
	The ordered Lyapunov spectrum showing the top 6 Lyapunov exponents are all negative, indicating asymptotically vanishing gradients.
	\textbf{(B)}
	Two stable fixed points representing a multistable system (Sec.~\ref{sec:decouplingmemory}).
	The neural activity of two neurons (color) over multiple trials visualizes their bistability.
	The adjoint signal for multiple trials show that learning signal decays to zero (in backward time).
	\textbf{(C)}
	A line attractor (pink line) representing continuous attractor dynamics (Sec.~\ref{sec:cattractor}).
	Each neurons activity converges over time.
	The neural activity of one neuron over multiple trials shows convergence to a continuum of values that encodes a gradation of initial conditions.
	The adjoint signal for a single neuron over multiple trials is constant and non-zero.
	The maximum Lyapunov exponent is zero, indicated by the $\dagger$ symbol, supporting one dimension of non-exploding, non-vanishing gradient propagation over time.
	\textbf{(D)}
	A randomly connected recurrent network with large gain representing chaotic dynamics.
	The neural activity of a subset of neurons over time show complex, seemingly random behavior.
	Small perturbations to the initial activity lead to very different trajectories over time.
	The adjoint signal quickly gets out of hand (backwards in time).
	There are 4 positive Lyapunov exponents, indicating the presence of asymptotically exploding gradients.
	\textbf{(E)}
	A stable limit cycle dynamics (Sec.~\ref{sec:pta}).
	At least 6 neurons participate in the limit cycle.
	Both the neural activity and adjoint signal are periodic.
	There is one null Lyapunov exponent corresponding to the oscillatory phase variable.
    }
    \label{fig:ALE_summary}
\end{figure}

In general, for an $n$ dimensional dynamical system, there exist $n$ independent Lyapunov exponents, whose collection is called the Lyapunov spectrum~\cite{Barreira2006}.
For Lyapunov-regular attractor dynamics, the Lyapunov spectrum is independent of the initial condition within the same basin of attraction by the Oseledets theorem~\cite{Barreira2006,Nayfeh1995}.
Therefore, the asymptotic EVGP is characterized by the Lyapunov spectrum corresponding to the basin of attraction that the trajectories of interest live in (Fig.~\ref{fig:ALE_summary}).
Under mild conditions, the Lyapunov spectrum, especially its sign, is invariant to diffeomorphism, therefore, only the topological structure of the dynamical system matters.

For each zero Lyapunov exponent, there is a corresponding dimension that carries the learning signal over time without exploding or vanishing at least exponentially fast.
Therefore, of great importance is the total number of zero Lyapunov exponents in the spectrum, to secure as many independent directions where the learning signals exhibit subexponential or bounded behavior.
Moreover, exploding gradients can induce sudden disruptive changes in the recurrent network and amplifies stochasticity in the learning signals, therefore deemed more harmful than vanishing gradients for learning.
To avoid EVGP, we seek topological structures of dynamical systems that \textbf{maximize the number of zero Lyapunov exponents while having no positive Lyapunov exponent}.
Similar non-asymptotic conditions have been argued for the well-posedness of deep neural networks~\cite{Haber2017,Hanin2018} and artificial recurrent networks~\cite{Chang2019}.

\subsection{Continuous attractor supports persistent memory and sensitivity}\label{sec:cattractor}
Continuous attractors are characterized by an attracting manifold such as a line, a ring, or a plane, where there is no flow, i.e., $\RN{\vx}{t} = 0$.
That is, small perturbations away from the manifold asymptotically returns to the manifold, and on the manifold neural activation patterns are persistent.
In the context of recurrent networks, the manifold of a continuous attractor is typically low-dimensional and embedded in a higher-dimensional neural activity space.
Since there is a continuum of stable equilibria where the neural activity does not change over time, the observed autonomous dynamics of the system are similar to a point attractor system, in that it generally decays to a fixed state and maintains a constant activity over time.
However, unlike point attractors, all states on the manifold exhibit similar behavior---perturbations on the manifold do not return to the initial state and hence are not forgotten.

As a conceptual tool in theoretical neuroscience, continuous attractors are widely used when working memory of continuous values is needed~\cite{Dayan2001}.
In combination with input, continuous attractors are also called integrators that are hypothesized to be the underlying computation for the maintenance of eye positions, heading direction, self-location, target location, sensory evidence, working memory, and decision variables, to name a few~\cite{Seung1996,Seung2000}.
A key signature of a recurrent network implementing a continuous attractor is the persistent neural activity that can be maintained at various levels during a memory or delay period~\cite{Romo1999}.
In neuroscience, a typical implementation of a continuous attractor are bump attractor network models where recurrent dynamics supports a particular bump of neural activation patterns to be self-sustained~\cite{Skaggs1995,Camperi1998,Renart2003,Noorman2022}.

Continuous attractors are considered biologically plausible, theoretically elegant, consistent with some neural recordings, and avoid asymptotic EVGP.
Given an $d$-dimensional continuous attractor manifold embedded within a recurrent dynamics of $n$-dimensions, it supports persistent continuous memory of $d$-dimension.
There are $d$ zero Lyapunov exponents corresponding to the perturbations tangent to the manifold, coinciding with the memory representation, and $(n-d)$ negative Lyapunov exponents that express the attractive nature.
In theory, the topology of the manifold can be arbitrary, ideally matching the desired structure of the target variables; for example, the ring attractor is natural for representing circular variable such as head direction.

\subsection{Structural stability and robustness of attractors}\label{sec:ss}
While continuous attractors have both persistent memory and persistent learning signals, their main weakness lies in their theoretical brittleness.
To understand how they can easily break down, let us first establish a conceptual framework on the nuance of noise.

Noise, practically defined as unpredictable components of the system's behavior, comes from many sources.
Based on the substance expressing noise in recurrent networks, we classify noise into two distinct types, corresponding to the two internal variables in \eqref{eq:ode}, the instantaneous neural activity vector $\vx(t)$ and the parameters of the underlying dynamical law $\vw$.
Noise in the neural activity and internal representations can be caused by noisy components and variability in the external input~\cite{Averbeck2006}.
We refer to this type of unpredictability injected into the neural state as the \emph{S-type noise}.
S-type noise encapsulates reversible changes in the neural state such that the deterministic part of the dynamics itself remains unchanged.

In addition to the S-type noise, biological neural systems have constantly fluctuating synaptic weights~\cite{Fauth2019,Shimizu2021}.
In other words, there is noise, which we call the \emph{D-type noise}, in the space of recurrent network dynamics parameterized by the synaptic weights.
Both in biological and artificial recurrent networks, a major source of D-type noise is learning.
Biological agents are constantly learning to improve performance, learn new tasks, and adapt to changing environmental structures and demands.
In machine learning, gradient-based learning is often implemented in an online or mini-batch fashion, where each update of the parameters inherits the randomness in the training dataset, making the training a stochastic process.
D-type noise has a potential benefit for avoiding local optima and providing an implicit regularization for overparameterized networks~\cite{Smith2021}.

An important consequence of the presence of D-type noise is that the neural computation implemented by recurrent dynamics is constantly fluctuating.
Therefore, the desirable properties of the dynamical system that require precise weight combinations are not stably achievable due to their unreliability.
In dynamical systems, when infinitely small changes in the parameters cause a qualitative change in the dynamics due to changes in the topology of the recurrent dynamics, it is said that the system goes under a bifurcation.
The topological structures that are robust under D-type noise are called \emph{structurally stable} -- for example, a stable fixed point is structurally stable~\cite{Kuznetsov1995}. 

There are three consequences of structurally stable components of a dynamical system.
First, neural computation implemented using structurally stable components is robust to the presence of D-type noise.
Second, structurally stable features are easier to learn via small changes and to maintain since once learned, it does not easily disappear due to D-type noise or further learning.
Finally, the asymptotic properties of the dynamics, such as their Lyapunov spectrum also do not suddenly change~\cite{Doya1992}.

Unfortunately, continuous attractors are not structurally stable---small changes in the dynamical system can easily destroy continuous attractors, and as a consequence, the corresponding Lyapunov exponent(s) moves away from zero.
For example, in machine learning, vanilla RNNs are sometimes initialized at the continuous attractor regime with all zeros such that $\RN{\vx}{t} = \tanh(\mathbf{0} \cdot \vx) = \mathbf{0}$, which avoids the asymptotic EVGP initially, but immediately loses the continuous attractor as a consequence of gradient-based learning.
This is a well-known problem in neuroscience, often referred to as the ``fine-tuning problem'' of the continuous attractor~\cite{Seung1996,Renart2003,Noorman2022}.
There have been remedies to lessen the degradation, often focusing on keeping the short-term behavior close to the continuous attractor case~\cite{Lim2012,Lim2013,Boerlin2013,Koulakov2002,Gu2022}.

The structural stability of the dynamical system implemented by a recurrent network critically depends on the architecture or the allowed parameters for the RNN.
It is possible to remove the brittleness inherent in continuous attractors by making the dynamics less flexible, for example, by requiring that some parameters are not optimized (i.e., they are not learnable parameters).
In machine learning, the so-called long short-term memory (LSTM) units are designed to withstand degradation by building in an independent line attractor per LSTM unit~\cite{Hochreiter1997}.
The continuous attractor in the original LSTM without the forget gate intentionally does not have any parameter that can induce its bifurcation (or disappearance).
However, in modeling biological systems, such constraints are not present in the theoretical and computational models of continuous attractors.

\subsection{Stable limit cycles and quasi-periodic toroidal attractors solve the EVGP}\label{sec:pta}
\begin{figure}[tbhp]
\centering
\includegraphics[width=5in]{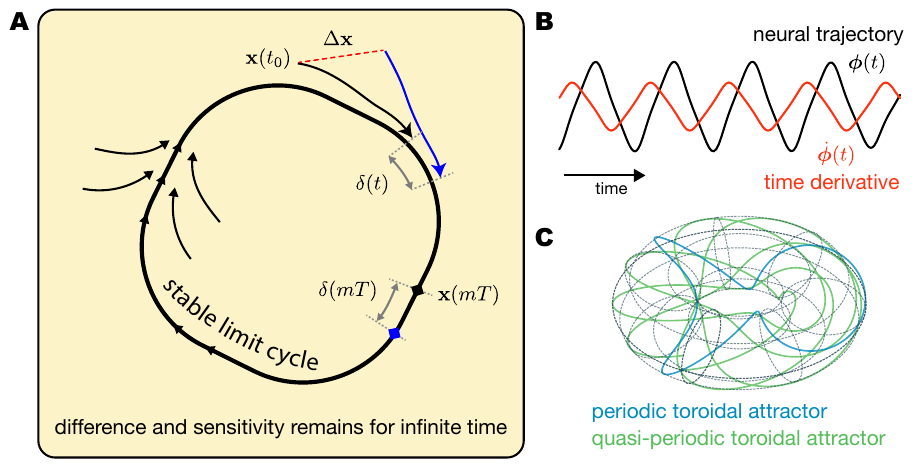}
\caption{
    \textbf{(A)} \textbf{Stable limit cycle dynamics do not forget phase differences.}
    Trajectories asymptotically converge to the limit cycle.
    On the limit cycle, the difference $\delta(t)$ is maintained and is periodic such that $\delta(mT)$ are the same for any sufficiently integer $m$.
    \textbf{(B)}
	Example neural trajectory and its time derivative.
    \textbf{(C)}
	Mathematically, two independent limit cycles may form a periodic or quasi-periodic toroidal manifold depending on their periods.
}
\label{fig:slc}
\end{figure}
Since continuous attractors are not suitable for learning long-temporal dependencies through gradient-based learning, we turn our attention to another well known system with zero Lyapunov exponents: stable limit cycles~\cite{Sokol2019a,Haken1983,Ermentrout2010}.
Stable limit cycle is composed of an attracting ring manifold, similar to the continuous attractor with ring topology.
But unlike the ring attractor, the neural activation is not persistent over time, but rather forms a periodic trajectory $\vphi(t)$ (Fig.~\ref{fig:slc}A).
The Lyapunov spectrum can be computed from the asymptotic trajectory $\vphi(t)$ with period $T$ known as the Floquet theory~\cite{Chicone2006}.
First note that the Jacobian matrix \eqref{eq:lyapunov:jacobian} as a function of $\vphi(t)$ is also periodic in $T$.
Therefore, the learning signals evolve in time with a periodically time-varying linear dynamical system \eqref{eq:variational}:
\begin{align}\label{eq:slc:sensitivity}
    \RN{\vDelta(t)}{t} = \vA(t) \vDelta(t),
\end{align}
where $\vA(t) = \vA(t+T)$ is the Jacobian.
Since $\vphi(t)$ is solution to the differential equation \eqref{eq:ode}, its temporal derivative $\dot{\vphi}$ (Fig.~\ref{fig:slc}B) also satisfies \eqref{eq:slc:sensitivity}:
\begin{align}
& \dot{\vphi}(t) \coloneqq \RN{\vphi}{t} = \vf(\vphi(t))
\\
\implies & \RN{\dot{\vphi}}{t} = \DF(\vphi(t)) \RN{\vphi}{t} = \vA(t) \dot{\vphi}
\end{align}
Therefore, any scalar multiple of $\dot{\phi}$ is a solution to \eqref{eq:slc:sensitivity}, therefore an additive part of the sensitivity over time.
In other words, as the network state approaches the stable limit cycle attractor, the corresponding adjoint and sensitivity signals do not decay nor explode but rather becomes periodic, giving rise to a zero Lyapunov exponent.

The zero Lyapunov exponent can be seen as preserving the asymptotic phase difference.
Any perturbation corresponding to the forward or backward flow of time on the limit cycle (i.e. tangent to the limit cycle manifold, see Fig.~\ref{fig:slc}A) would change the (asymptotic) phase and would be maintained~\cite{Barreira2006}.
The remaining Lyapunov exponents are negative due to the attracting nature.
For a real-valued phase space, at least two dimensions (or neurons) are needed to sustain one zero-Lyapunov exponent~\cite{Strogatz2000}.
Note that the representation of the perturbation and the accumulated gradient are necessarily circular in topology.

We can increase the number of zero Lyapunov exponents of the system by introducing independent limit cycles.
In the simplest case, the stable limit cycle dynamics would occupy separate subspaces, for example, each pair of neurons only partakes in one limit cycle dynamics.
The joint representation in this case is bound to be toroidal (Fig.~\ref{fig:slc}C).
If the periods of limit cycles are rational multiples of each other, the joint neural state is periodic.
Therefore, we call the resulting dynamical system the \emph{periodic toroidal attractor (PTA)}.
When the periods of the independent limit cycles are not mutually rational multiples of each other, the orbit becomes \emph{quasi-periodic}---it never repeats itself in the joint neural state space, and every neural trajectory densely fills the torus.
We call this case the \emph{quasi-periodic toroidal attractor}.
Both cases, however, the maximum number of zero LEs is $n/2$, hence we refer to both cases as the PTA for simplicity.

An important characteristic of stable limit cycles and PTA is that they are structurally stable~\cite{Guckenheimer1983}.
Unlike generic continuous attractors, independent of the parameterization of the dynamical system, D-type noise does not easily destroy the topological structure that provides the solutions to asymptotic EVGP.
It also implies that stable limit cycles may not be rare.
In fact, nonlinear oscillations in various neural models naturally lead to stable limit cycle solutions, and their signatures are ubiquitously observed in biological dynamical systems~\cite{Buzsaki2023}.
We conjecture that PTAs are the only structurally stable dynamics in general that support persistent learning signal over time.
We summarize the theory of recurrent dynamics and learning-signal behavior in Fig.~\ref{fig:ALE_summary}.

\subsection{Toroidal attractor initialization experiments}
\begin{figure}[tbp]
  \centering
  \includegraphics[width=0.7\textwidth]{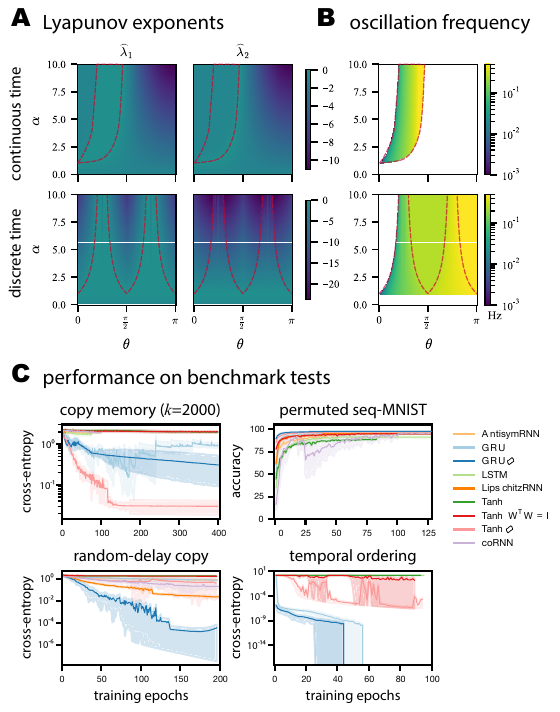}
  \caption{
	\textbf{Numerical experiments using artificial neural networks.}
	\textbf{(A)}
	    Numerically evaluated Lyapunov spectrum ($\hat{\lambda}_1 \geq \hat{\lambda}_2$) as a function of rotation speed $\theta_i$ and amplitude gain $\alpha_i$ parameters in~\eqref{eq:blockortho}.
	    Dashed lines indicate the region where the leading Lyapunov exponent $\hat{\lambda}_1$ is zero.
	    The continuous time system and the discrete time system behaves similarly but with different parameters.
	\textbf{(B)}
	    Region of stable limit cycle detection by the spontaneous generation of periodic trajectories.
	\textbf{(C)}
	    A barrage of difficult benchmark tests involving long range temporal or sequence dependence learning.
	    Test cross-entropy loss or accuracy during training shows faster convergence and also finding better performing models for PTA GRU (\lcglyph) and PTA tanh (\lcglyph) networks.
      }
    \label{fig:rnn:experiments}
\end{figure}

To test the practical efficacy of PTA, we performed a barrage of experiments where we trained artificial recurrent neural networks (RNNs) using backpropagation through time (BPTT)~\cite{Sokol2023b}.
For concreteness, we choose vanilla ($\tanh$) and gated recurrent unit (GRU)~\citep{Jordan2019a} architectures but note that our theory is completely agnostic to the particular architecture choices.
For both architectures, to obtain a stable limit cycle, we parameterized the recurrent connectivity weight matrix with a scaled rotation matrix~\citep{Beer1995,Jordan2019a}.
To easily form a collection of independent oscillators, we impose the follow block matrix structure:
\newcommand{\Winit}{\mathbf{W}_\text{init}}
\begin{align}\label{eq:blockortho}
\Winit &=
    \begin{bmatrix}
	\alpha_1\left(
	\begin{matrix}
	    \cos(\theta_1) & -\sin(\theta_1)\\
	    \sin(\theta_1) & \cos(\theta_1)
	\end{matrix}\right) & & \mathbf{0}  \\
    & \ddots & \\
	\mathbf{0} & &
	\alpha_m \left(\begin{matrix}
	    \cos(\theta_m) & -\sin(\theta_m)\\
	    \sin(\theta_m) & \cos(\theta_m)
	\end{matrix}\right)
    \end{bmatrix}
\end{align}
Note that $\Winit$ has a block orthogonal structure.
For each of the $2 \times 2$ block, depending on the $(\alpha_i, \theta_i)$, the pair of coupled neurons can exhibit spontaneous oscillations.
Fig.~\ref{fig:rnn:experiments}A shows the Lyapunov spectrum as a function of the two parameters for each of the $2\times 2$ block of either continuous or discrete time formulations; For example, for vanilla RNNs, it analyzes the following spontaneous forms:
\begin{align}\label{eq:cvd:tanh}
    \dot{\vx} &=
	-\vx +
	\tanh\left(
		\Winit \cdot \vx
	\right)
	&\qquad \text{(continuous time)}
    \\
    \vx(t+1) &=
	\tanh\left(
		\Winit \cdot \vx(t)
	\right)
	 &\qquad \text{(discrete time)}
\end{align}
Generally, and strong enough oscillatory gain $\theta_i$ is needed for a specific amplitude gain $\alpha_i$ to produce an oscillation (Fig.~\ref{fig:rnn:experiments}B).
For the numerical experiments, we uniformly sample from the identified parameter regions corresponding to having a zero Lyapunov exponent, thus forming a pool of nonlinear oscillators with different frequencies before training.
As we trained the RNNs, we did not constrain the dynamics to the block orthogonal structure of \eqref{eq:blockortho}, but let any neuron-to-neuron connection in the recurrent weights to become non-zero.

To evaluate the efficacy of PTA as an initialization strategy (denoted by \lcglyph), we conduct experiments on a host of standard benchmarks and compare the results with models tailored to learn long--range temporal dependencies, such as the AntisymmetricRNN, coRNN and the LipschitzRNN~\citep{Chang2019,Rusch2021,Erichson2021}.
As additional baselines, we include vanilla $\tanh$, GRU, and LSTM architectures.
Fig.~\ref{fig:rnn:experiments}C summarizes the learning on the standard benchmark tasks.
Details of the experiments and additional experimental results can be found in~\ref{sec:appendix:RNNexp}.
We matched the size of the network and optimized hyperparameters for all RNNs for a fair comparison.
For the copy memory and sequential CIFAR tasks, only the PTA initialized networks were able to reliably learn the tasks.
In the other two benchmarks, PTA initialized RNNs learned faster and finds better solutions.
Although the initial dynamics were oscillatory before training, the gradient-based learning often reaches solutions without oscillation (data not shown).

Note that the for the vanilla and GRU RNNs, the standard initializations and the PTA block orthogonal initialization share the same space of potential solutions.
The difference is the learnability---some initial conditions are not suitable for reaching parts of the solution space with (stochastic) gradient descent.
The PTA initialization provides useful gradients across long temporal gaps required to start moving to the right direction in the solution space.
We trained the RNNs with BPTT which is a non-biologically plausible machine learning technique to demonstrate even with powerful learning algorithms, the existence and robustness of zero Lyapunov exponents of the recurrent dynamics are beneficial.
We discuss the biological plausibility in Sec.~\ref{sec:biorelevance}.

\subsection{Implications on biological temporal learning}
Although the exact biophysical mechanism of gradual learning is yet to be identified, we have conjectured that \mbox{(quasi-)}periodic attractors are theoretically necessary for carrying the gradient information needed for learning long temporal relations.
If we take this mathematical implication at face value, it suggests that the neural system that carries the learning signal over time should contain stable oscillations.
There are many potential biophysical mechanisms that can produce stable oscillations, from periodic bursting neurons, subthreshold oscillations in gap-junction coupled inferior olivary network, corticothalamic loops, interhemispheric mutually excitatory feedback loops, Wilson-Cowan model, KII set, to cortical E-I dipole oscillators~\cite{Wilson1972,Freeman1975,Izhikevich2007}.
Note that any biophysical mechanism that can generate stable oscillation can only do so with recurrent dynamics, but not necessarily at the neuronal network level.
We assume that the observed neural activity from various modalities (electrophysiology, imaging, and so on) reflects the internal neural state evolution over time, and hence the temporal learning signals, albeit only partially.
Following is a list of implications if the biological system utilizes the learning signal over time carried by PTA:

\begin{enumerate}[label=\textbf{L\arabic*}.]
    \item \textbf{Neural activity are spontaneously periodic/quasi-periodic and stable in amplitude}
	Nonlinear oscillations are plentifully observed throughout the central neural system, especially when the animal is ``idle''.
	These oscillations are not completely synchronous, and the power spectrum for each oscillation band shows a broad bump consistent with PTA formed from a distributed range of periods.
	Given a perturbation, whether they are sensory, internally generated, or direct manipulation in nature, the amplitude of the oscillations may briefly change but relax back to the baseline level in the absence of further perturbations.

    \item \textbf{Learning signals are (quasi-)periodic in time.}
	Generalized eligibility traces that correspond to parameters that are modified in a learning paradigm to solve the temporal credit assignment problem, oscillate in time.
	They may be present within a single neuron as in synaptic or dendritic states, but also at a micro-circuit or neuropeptide scale.

    \item \textbf{Paradoxical timing dependent anti-learning.}
	Due to the circular or toroidal topology, the accumulated learning signals in one direction eventually become similar to learning signals in the opposite direction, that is, learning signals are \mbox{(quasi-)}periodic in phase space.
	Therefore, there could be certain delays that may increase the error in the updated behavior.
	Such paradoxical delay-dependent learning effect could be measured from the evolution of behavior during the learning process.
	More generally, if the presentation of the stimulus is time-locked to the relevant oscillations, certain timing relations may be easier or more difficult to learn due to non-homogeneous nonlinear effects.

    \item \textbf{Successful learning of long temporal dependency may weaken the spontaneous oscillations.}
	Learning a task that does not require precise timing nor periodic output could reduce the reservoir of oscillators as the solution lacks oscillations.
	As the animal becomes better trained, spontaneous oscillations may bifurcate away, at least until they are replenished through yet unknown homeostatic mechanisms.

    \item \textbf{The effectiveness of learning shows a U-shape modulation with respect to the strength of oscillations.}
	Our theory predicts that long-temporal dependency learning is more difficult in the absence of oscillations.
	On the other hand, the number of null Lyapunov exponents in a collection of stable limit cycles collapses to one when the frequency of the oscillators perfectly synchronize and remain phase-locked.
	In either regime, in the absence of oscillations or when the full population is in low-D synchronized dynamics, learning signals cannot temporally propagate using PTA.
	Rather, the number of null Lyapunov exponents is maximized in the intermediate level, when the neural oscillators are weakly coupled or independent.
	Therefore effective learning of temporal learning requires (perturbation-dependent) desynchronization, which will result in weaker strength of observed oscillations, for example, in the field potentials.
	Experimental validation could be in the form of an opportunistic analysis or closed-loop experiments of learning on the strength of relevant ongoing oscillations.
\end{enumerate}

\subsection{Persistent memory with quasi-periodic toroidal attractors}\label{sec:relphasememory}
In section~\ref{sec:decouplingmemory}, we saw that recurrent dynamics with long-lasting short-term/working memory can have poor learning signals.
Toroidal periodic attractors have persistent learning signals and are robust to D-type noise, but do they have long-lasting short-term memory properties?
As we have seen in section~\ref{sec:pta}, the perturbation in the phase is maintained in time, hence we have \emph{relative phase memory}.
But to read out the phase difference, we need to know the counterfactual phase without the perturbation.
One potential solution is to have a copy of the oscillator that keeps track of the unperturbed phase, such that the phase difference can be read out in a phase-independent manner. 

Unfortunately, having an identical copy is sensitive to parameter tuning, i.e., the two identical nonlinear oscillators' solution is not structurally stable.
If parameters that govern the period of one of the oscillators is perturbed, the phase difference is no longer constant over time.
This is essentially the same problem as the fine-tuning problem that plagues continuous attractors.

Luckily, there are biophysical constraints that can counter the mathematical structural instability of relative phase memory.
Temporal dynamics that are \emph{reliable and not adjustable} such as intracellular biochemical processes, conduction delays, and membrane and synaptic time constants could form a basis for structural stability.
As we will discuss in the next section (Sec.~\ref{sec:eb}), for example, a neuron reciprocally connected to a neuron in the contralateral hemisphere~\cite{Deutsch2020} can form an oscillator unit through the feedback loop.
The constant conduction delay and physical distance of the feedback loop make the period of oscillation fixed.
Moreover, neurons that share the same reciprocal pathways will have near-identical periods.
Therefore, quasi-periodic toroidal attractors with phase-independent readout mechanisms could be behind biological persistent activities and short-term memory.

If the contents of working memory is encoded in relative phases of oscillators, we can make several predictions:
\begin{enumerate}[label=\textbf{M\arabic*}.]
    \item Contents of working memory should be decodable from single trial phase code~\citep{Maris2013,Agarwal2014}.
    \item Resetting the oscillation phases would instantly erase the contents of working memory. 
    \item Potentially useful stimulus induces phase shifts of corresponding neural oscillations.
    \item Input-driven desynchronization may be observed in neural circuits where corresponding working memory is needed.
    \item If quasi-periodic oscillation underlies persistently \emph{constant} activity, lesions of the readout circuit will not fundamentally destroy the memory.
	If the readout has multiple dimensions, the intact neurons may continue to show persistent activity.
	Continuous attractor dynamics and neural integration, on the other hand, will be completely disrupted if a significant portion of persistent activity neurons are destroyed.
\end{enumerate}

\subsection{Persistent activity bump on a ring without continuous attractors}\label{sec:eb}
\newcommand{\PRm}{\ensuremath{\textrm{P}^{-}_{\textrm{R}}}}
\newcommand{\PRp}{\ensuremath{\textrm{P}^{+}_{\textrm{R}}}}
\newcommand{\PLm}{\ensuremath{\textrm{P}^{-}_{\textrm{L}}}}
\newcommand{\PLp}{\ensuremath{\textrm{P}^{+}_{\textrm{L}}}}
\begin{figure}[tbhp]
    \centering
    \includegraphics[width=\textwidth]{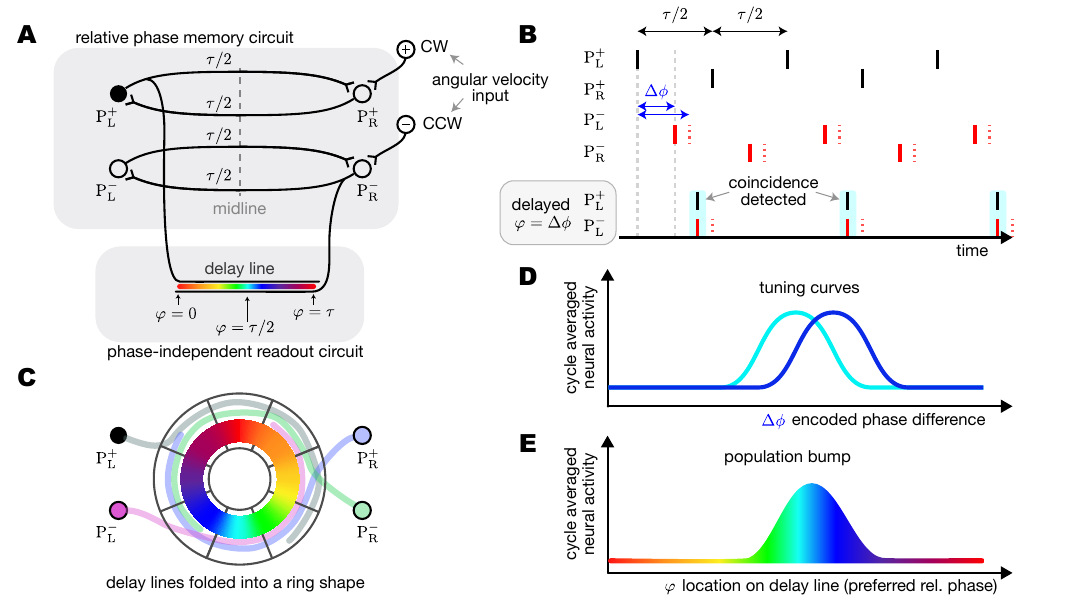}
    \caption{
	\textbf{Head direction memory with a robust bump on a ring but implemented with two pairs of oscillators.}
	\textbf{(A)}
	    Two intrinsically oscillating neurons $\PRp$ and $\PLp$ are coupled across the hemisphere with $\tau/2$ delay, forming a stable $\tau$-periodic oscillator.
	    Independently $\PRm$ and $\PLm$ form an identical oscillator circuit.
	    Clockwise velocity input is integrated into the phase of the $+$ oscillators and vice versa (only half of the circuit depicted).
	    The relative phase between the $+$ and $-$ oscillators $\Delta \phi$ encodes the head direction.
	    With an appropriate delay through axon length, phase differences corresponds to coincidence at different location.
	\textbf{(B)}
	    $\tau$-periodic activity of each of the 4 neurons in the oscillator.
	    Red lines and red dashed lines correspond to two different head direction memories.
	    Note that a continuum of phase differences can be represented.
	    Given a particular phase difference $\Delta\phi$, appropriately delayed spike trains coincide periodically.
	\textbf{(C)}
	    The delay line in (A) is better formed into a 2-dimensional anatomical structure.
	    Note that the $+$ oscillator neurons project in CW direction, while the $-$ oscillators wrap around in CCW direction.
	    The underlying 8-segments that corresponds to the ellipsoid body of the Drosophila central complex.
	\textbf{(D)}
	    The interaction between delayed periodic neural activity arriving at each location on the ring has a tuning curve with respect to $\Delta\phi$.
	    The response is periodic in $\tau$, and a low-pass filter can average out the periodic fluctuation.
	\textbf{(E)}
	    At a time scale of smoothed responses, the population activity over the ring manifests as an activity bump.
	    This bump is indistinguishable from the continuous bump attractor at this time scale.
	    Unlike the bump attractor model, the proposed mechanism is robust to D-type perturbations.
    }
    \label{fig:relativeEBcircuits}
\end{figure}

The ellipsoid body, or the so-called `compass' system, in the Drosophila brain has an anatomical shape of a ring, and exhibits a stable bump of activity corresponding to the animal's head direction in a persistent and continuous manner~\cite{Seelig2015,Kim2017}.
Since these properties are hallmarks of the ring attractor model---a continuous attractor with a circular topology---several models have been proposed to capture the essence of the underlying biophysical recurrent dynamics~\cite{Kakaria2017,Kim2017,Heinze2018,TurnerEvans2017,TurnerEvans2020,Noorman2022}.
Initial models required precise symmetric connectivity and the mean-field limit of infinite neurons to form the continuous translation of the bump on the ring~\citep{Zhang1996}.
Later models were able to achieve it with a finite size implementation~\citep{Noorman2022}, and investigated heterogeneous connectivity~\citep{Darshan2022}.
Nevertheless, continuous attractor networks cannot be exactly implemented by a biological network, because of their structural instability (Sec.~\ref{sec:ss}).
In the following, we present an alternative mechanism that implements a continuous and persistent head-direction representation with a phase-independent PTA memory.

We hypothesize that there are two independent oscillators (denoted by `$+$' and `$-$') with identical periods.
In particular, each of the oscillators are implemented with a pair of spiking neurons, one on each hemisphere (Fig.~\ref{fig:relativeEBcircuits}A).
There are many possible mechanisms for the pair to synchronize~\citep{Ermentrout2010};
For simplicity, we assume they are mutually excitatory, and antiphase locked such that the periods of each neuron is $\tau$, and the contralateral spikes are $\tau/2$ delayed (Fig.~\ref{fig:relativeEBcircuits}B).
Given the clockwise (`$+$') component of the angular velocity of the animal's heading over time, $V^{+}_a(t) = \left[V_a(t)\right]^+$, the temporal integration can be folded into the phase of oscillation corresponding to \PLp\ (and \PRp\ with $\tau/2$ offset):
\begin{align}
    \phi^{+}(t) &= \frac{\tau}{2\pi} \int^t V^{+}_a(s) \mathrm{d}s + t \mod \tau
\end{align}
where the modulus operator wraps the phase to be represented in units of time between $0$ and $\tau$.
We can define $\phi^{-}(t)$ similarly for the $\left(\PLm, \PRm\right)$ pair that integrates counter-clockwise (`$-$') angular speed $V^{-}_a(t) = -\left[-V_a(t)\right]^+$.
\begin{align}
    \phi^{-}(t) &= \frac{\tau}{2\pi} \int^t -V^{-}_a(s) \mathrm{d}s + t \mod \tau
\end{align}
These integration of angular velocity as additive time delay in the phase can be implemented with synaptic input to the appropriate neurons (Fig.~\ref{fig:relativeEBcircuits}A, CW \& CCW) with a constant phase resetting curve~\citep{Ermentrout2010}.
We observe that \emph{the phase difference integrates the angular velocity and is otherwise independent of time}:
\begin{align}
    \Delta\phi(t) &= \phi^{+}(t) - \phi^{-}(t)
    \\
	&= \frac{\tau}{2\pi} \int^t \left(V^{+}_a(s) + V^{-}_a(s)\right) \mathrm{d}s \mod \tau
    \\
	&= \frac{\tau}{2\pi} \int^t V_a(s) \mathrm{d}s \mod \tau
\end{align}
Hence, $\Delta\phi(t)$ implements a relative phase memory that encodes the animal's heading direction.

Although the relative phase memory $\Delta\phi(t)$ is constant in time in the absence of angular velocity input,
each of the four neurons in the memory circuit exhibits a time varying oscillation, making access to $\Delta\phi$ non-trivial.
Inspired by the Jeffress model and the corresponding circuit in the brain stem that detects sub-millisecond phase difference necessary for sound localization~\citep{Feddersen1957,Mc_Laughlin2010,Carr1990}, we propose a similar axonal time delay line.
In Fig.~\ref{fig:relativeEBcircuits}A, one of several possible configurations that form a delay line such that the location along the axon action potentials from arrive coincidentally.
This maps the relative phase memory to a location of coincidence in a time/phase-independent manner (Fig.~\ref{fig:relativeEBcircuits}B, bottom).

There are at least two plausible maps from the abstract location of coincidence onto the circular anatomy of the ellipsoid body.
First is to project from the delay line structure (for example, formed by the intrinsic neurons in the protocerebral bridge) to the ellipsoid body~\citep{Kakaria2017}.
Second is to wrap the delay line around the ring as depicted in Fig.~\ref{fig:relativeEBcircuits}C.
However, in the latter configuration, it is required that the two oscillators wrap around the ring are predicted to be in opposite orientations which is asymmetric and  developmentally less plausible.

Suppose we have a dense distribution of synaptic connections along the delay line.
The synaptic activity can be modeled with a tuning curve as a function of three variables,
\begin{align}\label{eq:eb:tuning}
    f(t, \Delta\phi(t), \varphi)
	&= \Phi\left(h \ast \PRm(t + \Delta\phi(t) - \varphi), h \ast \PLp(t)\right)
\end{align}
where $t$ is time, $\varphi$ is the location on the delay line expressed in units of time delay, $h$ is a temporal filter that captures the synaptic and membrane dynamics that effectively smoothes the neural activity, $\ast$ denotes temporal convolution, and $\Phi(\cdot, \cdot)$ captures the nonlinear coincidence interactions.
The tuning curve $f$ is periodic in each of its three arguments:
$f(t, \cdot, \cdot)$ is periodic over time of $\tau$ in the absence of angular velocity,
$f(\cdot, \Delta\phi(t), \cdot)$ is periodic in $2\pi$ as the animal's heading makes a full rotation,
and $f(\cdot, \cdot, \varphi)$ should be periodic in $\tau$ to faithfully map $\Delta\phi(t)$ to location along the delay line.

We further assume that the readout and the measurement of the neural activity is low-pass filtered by temporal smoothness of signals and also the Ca$^{2+}$ fluorescence indicators.
With an appropriate low-pass filter, the periodic fluctuations are effectively averaged out.
Therefore, we can obtain the cycle average tuning curve,
\begin{align}\label{eq:eb:tuning:circ}
    \mathring{f}(\Delta\phi(t), \varphi) &= \int_{t-\tau}^t f(t, \Delta\phi(t), \varphi) \mathrm{d}t
\end{align}
which only depends on the relative phase memory and location along the delay line.

For a particular coincidence detector along the delay line, the cycle-averaged tuning curve with respect to the memory or integrated heading (Fig.~\ref{fig:relativeEBcircuits}D) is a symmetric bump of activity such that the maximum activity is reached when the physical delay matches the phase memory content.
At the same time, the population activity of all coincidence detectors along the delay line also forms a \emph{spatial} bump of activity
 as a function of $\varphi$ (Fig.~\ref{fig:relativeEBcircuits}E), where the shape of the bump is determined by $\Phi$ and $h$.
Since \eqref{eq:eb:tuning} only depends on the difference $(\Delta\phi - \varphi)$, the two bumps are of the same shape.
The spatial coincidence detector activity pattern is consistent with the observations in the ellipsoid body, however, this model avoids the fine-tuning problem of continuous attractors.
Moreover, neurons corresponding to the 8 segments (Fig.~\ref{fig:relativeEBcircuits}C) of the ellipsoid body can spatially integrate this information to produce stable 8-dimensional activity bump.

\section{Discussion}
Fading memory, multiple stable fixed points, and continuous attractor networks have been the dominant theoretical frameworks for modeling the neural dynamics underlying long-term learning and short-term/working memory.
In this exposition, we analyzed the strengths and weaknesses of recurrent dynamical systems on their ability to propagate gradient signals over time, and their robustness under D-type perturbations.
Our theoretical investigation revealed that temporal propagation of learning signals is not necessarily a consequence of persistent memory, but rather the topology of the dynamical system.
In the asymptotic limit of temporal extent, the Lyapunov spectrum summarizes the behavior of learning and memory:
negative Lyapunov exponents correspond to vanishing gradient signals, while zero Lyapunov exponents correspond to persistent working memory subspace and learning signals.
Our theory can be generally applied to \emph{any} continuous dynamical system, irrespective of the biophysical details or the specific architectures of the system: it equally applies to a single synapse, a single neuron, a network of neurons, or heavily engineered complex gated recurrent neural networks.

We showed that continuous attractors are not necessary to implement continuous working memory and not suitable for learning long-temporal dependency learning.
For attractor dynamics, the number of zero Lyapunov exponents determines the ideal configuration for learning and continuous-valued memory.
We conjecture that there are only two typical dynamical systems that achieve a reasonable solution to the requirement of persistent memory and learning: the continuous attractor networks and the lesser-known (quasi-)periodic toroidal attractors (PTA).
Unlike the continuous attractors, which quickly succumb to D-type noises such as continuous synaptic rewiring~\citep{Loewenstein2015}, PTAs are robust against D-type noise and can also bridge arbitrarily long intervals of time.
The simplest PTA is the stable nonlinear oscillator which has the same attractor manifold as a subtype of continuous attractor, namely the ring attractor.
However, they do not share the same vector field topology, since the stable limit cycle has no stationary points while the ring attractor has a continuum of them.

Our theory brings a fresh perspective to learning and memory in theoretical neuroscience and machine learning.
We showed that RNNs initialized to exhibit PTA are better at learning long-range dependencies.
We predict that the spontaneous nonlinear oscillations in the biological systems may partake in carrying the learning signal and memory content, and modulation of the oscillations may disrupt learning and working memory.
We present a novel implementation of the neural integration of angular velocity using PTA and phase-independent readout mechanism that relies on delay-lines that recapitulates the observations of the activity bump in the ellipsoid body.

\subsection{Relevance of asymptotic analysis}
Our analysis focuses on the asymptotic time horizon, while behavior and learning occur on finite time scales in practice.
The asymptotic analysis, albeit an approximation of the finite time scales, can provide practical insights since typical dynamical systems at rest converge (or diverge) exponentially fast near the attractor without a long transient behavior.

\subsection{Relevance of spontaneous attractor dynamics}
We assumed the network operates in the weakly input driven regime, that is, the asymptotic sensitivity analysis is dominated by that of the autonomous (spontaneous) dynamics.
For example, a group of uncoupled leaky integrate and fire neurons would collectively converge to the their resting membrane potentials, and for sufficiently weak input, fluctuate around the resting potential.
Furthermore, gating and attention mechanisms that have been successful in machine learning indicates that protecting the spontaneous dynamics by limiting input drive is a viable strategy that biology may be incorporating.
However, the theory in the current form is limited for analyzing strongly input driven recurrent networks---the local Lyapunov spectrum can be analyzed, but the global properties based on topology are intractable.

\subsection{Generalized gradient descent and differentiability}\label{sec:gengrad}
Although we focused our discussion on gradient descent where throughout learning, the parameters are updated in a way that optimally minimizes the loss function,
any method that learns \emph{incrementally} will inherently behave similarly as the most efficient path to the minimum will always align with the gradient~\citep{Richards2023}.
Consequently, any first-order learning algorithm can only match, but not surpass, the local optimization capabilities of gradient descent.
This naturally establishes gradient descent as an invaluable yardstick in the landscape of learning algorithms.

It is important to note that there exist different notions of differentiability, and these allow us to take derivatives even in the context of spiking neural networks.
Mathematically, these networks are often considered non-differentiable due to their binary nature and threshold firing mechanisms.
However, recent advances such as surrogate gradients and EventProp~\citep{Wunderlich2021} demonstrate that suitable generalizations considerably broaden the class of problems that are amenable to analysis in terms of calculus.
Our theory predicts that spiking neural network in the spontaneous oscillation regime, e.g., having a bias current for each neuron to have quasi-periodic behavior at rest can improve gradient propagation through time~\cite{Huh2019}.

\subsection{Biological relevance of gradient based learning in recurrent networks}\label{sec:biorelevance}
Backpropagation has been the foundation of machine learning and adaptive signal processing, however it is not easily mapped to a biologically plausible learning rule~\cite{Lillicrap2020}.
Moreover, backpropagation through time (BPTT), a numerically efficient method for training recurrent networks in machine learning, has additional difficulties since physical computation of the gradients requires neural states to be accessed backwards in time---naively, it is not physically possible to implement it in a causal manner without additional memory mechanisms.
Instead of the backward adjoint dynamics~\eqref{eq:adjoint}, it is possible to use the forward sensitivity~\eqref{eq:sensitivity} dynamics for computing the necessary gradients, known as real-time recurrent learning (RTRL)~\cite{Williams1989,Marschall2019}.
RTRL produces gradients equivalent to BPTT in an online manner, but notoriously demanding in memory and computation in general:
It requires perfect, continuous-valued memory in the order of number of parameters times dimensionality of the state space, again not biologically plausible.
Approximate forms of RTRL that operates with biological recurrent networks is an active area of research~\cite{Murray2019,Roth2019,Bellec2020}.
Finding a biologically plausible way to implement BPTT for PTA is an open problem.

\subsection{Timing information in oscillations}
In addition to carrying persistent learning signals through time, oscillations have an advantage over continuous attractors.
It provides temporal basis functions that can carry timing information~\citep{Miall1989,Buhusi2005,Park2014a}.
Further investigations in implementation and learning of timing can yield insights into neural computation and we leave it for future work.

\subsection{Other working/short-term memory mechanisms that uses oscillations}
There are several prior work that utilized oscillations to study learning and working memory.
\citet{Lisman1998} proposed a bistable oscillation induced by NMDA-receptors.
\citet{Champion2023} proposed a multistable dynamics using discrete frequencies of oscillations.
\citet{Noest1988} proposed complex phasors to implement discrete attractor networks (see also~\cite{Frady2019}).
\citet{Orvieto2023} proposed an array of complex oscillators to produce stable linear dynamics.
In none of these works a continuous gradient propagation was analyzed or a continuous value working memory mechanism has been investigated.

Notably \citet{Burgess2008} proposed a continuous integration in the phase of oscillations in the context of spatial integration.
Their velocity-controlled oscillator represents the head-direction with a phase code similar to what we proposed in Sec.~\ref{sec:eb}, however, their mechanism is proposed to be contained within a single cell and focused on explaining the spatial periodicity of the grid field rather than a temporal learning signal or general theory of short-term memory.

\section*{Acknowledgements}
MP and PS were supported by NSF CAREER Award IIS-1845836 and NIH RF1-DA056404.
MP and AS were supported by Champalimaud Foundation.

\section*{Author contributions}
MP and PS conceptualized the project.
MP and PS designed the models and conducted formal analysis.
PS and AS ran the experiments.
MP wrote the original manuscript.
All authors reviewed and edited the manuscript.
MP, PS, and AS created the visualizations.

\section*{Declaration of interests}
MP is co-founder of RyvivyR Inc.

\bibliographystyle{plainnat_memming_v1}
\bibliography{refs,catniplab}

\appendix
\setcounter{section}{0}
\setcounter{equation}{0}
\renewcommand{\thesection}{S\arabic{section}}
\renewcommand{\thefigure}{S\arabic{figure}}
\renewcommand{\thetable}{S\arabic{table}}
\renewcommand{\theequation}{S\arabic{equation}}

\section*{Supplementary Material}
\section{Details for generating Fig.~\ref{fig:ALE_summary}}
The code to reproduce Fig.~\ref{fig:ALE_summary} can be found at \url{https://github.com/Asagodi/ALE}.

\subsection{Ordinary differential equations}
The ODEs for Fig.~\ref{fig:ALE_summary} were solved with explicit Runge-Kutta method of order 4.
For each network ten random initial values were sampled.
The adjoints were solved backwards in time with the same method.  One of the random initial values was used for the calculation of the adjoint.

The activity of the 2 fixed point and line attractor network systems is given by:
\begin{equation}
\RN{x_i}{t} = -x_i + \operatorname{ReLU}\left(\sum_{j=1}^NW_{ij}x_j+b_i\right)
\end{equation}
with connectivity matrix $\vW$ and bias term $\vb$.
The size of both networks is 6 units.
For the 2 fixed point network the connectivity matrix is given by
\[\vW=\begin{bmatrix}
.35 & -1 & 0 & 0 & 0 & 0 \\
-1 & .35 & 0 & 0 & 0 & 0 \\
0 & 0 & -0.01 & 0 & 0 & 0 \\
0 & 0 & 0 & -0.02 & 0 & 0 \\
0 & 0 & 0 & 0 & -0.03 & 0 \\
0 & 0 & 0 & 0 & 0 & -0.01
\end{bmatrix}.  \]
and for the line attractor network
\[\vW=\begin{bmatrix}
0 & -1 & 0 & 0 & 0 & 0 \\
-1 & 0 & 0 & 0 & 0 & 0 \\
0 & 0 & -2 & 0 & 0 & 0 \\
0 & 0 & 0 & -2 & 0 & 0 \\
0 & 0 & 0 & 0 & -2 & 0 \\
0 & 0 & 0 & 0 & 0 & -2
\end{bmatrix}.  \]
For both the 2 fixed point and line attractor network systems the bias term is given by $\vb=\textbf{1}$.

The activity of the chaotic and limit cycle networks are generated by the following ODE:
\begin{equation}
\tau\RN{x_i}{t} = -x_i +\sum_{j=1}^NW_{ij}\tanh(x_j),
\end{equation}
where $\tau= 10$ ms is the time constant.
The recurrent connectivity matrix is represented by the sparse $n \times n$ matrix $\vW$ which
has nonzero entries randomly chosen from a Gaussian distribution with zero mean and s.d., $g/ \sqrt{p_c n}$, where $g$ is the synaptic strength scaling coefficient, and $p_c = 0.1$ is the connection probability between units~\citep{laje2013}.
We used,
\begin{align}
    \vW_{ij} &= \begin{cases}
	\mathcal{N}(0,\frac{g^2}{p_c n}) & \text{with probability } p_c
	\\
	0 & \text{with probability } 1-p_c,
\end{cases}
\end{align}
with $g=1.8$.
For large networks, values of $g > 1$ generate increasingly complex and chaotic patterns of self-sustained activity. The size of the chaotic network was 100 units. The ten initial values for the numerical integration were generated from $\mathcal{N}(x_0, 10^{-8})$ where the reference point $x_0$ was chosen randomly.

The size of the limit cycle network was $n=20$ units with $g = 1.8$.
Random networks at this smaller size were frequently stable limit cycles and not chaotic. We picked one at random.
The ten initial values were generated from $\mathcal{N}(0, 1)$.

\subsection{Lyapunov exponents}
The Lyapunov exponents were approximated through a QR decomposition based method~\citep{Benettin1980} with the solved orbits.
For an orbit that converged to the invariant manifold evaluated at times $t_k=k\delta t$ for $k=1,\dots,t_1/\delta t$, set $R_0= \mathbb{I}$ and
$M_k = \mathbb{I} + x(t_k)\delta t$.
Then using the QR decomposition
$Q_{k+1}R_{k+1} = M_k Q_k$
average over the steps to get
$\lambda = \frac{1}{t_1 \delta t}\sum_{k=1} \log\left(\abs{R_{k,ii}}\right)$.

\section{Numerical BPTT experiments}\label{sec:appendix:RNNexp}
The `vanilla' variants of $\tanh$ RNNs and GRUs were initialized using the defaults provided in PyTorch~\citep{Paszke2019}, while the recurrent weights of orthogonally initialized $\tanh$ RNNs (denoted as $W\trp W = I$ for brevity) were sampled from the uniform measure over the orthogonal group~\citep{Vorontsov2017}.

 \subsection{Permuted seq-MNIST}
 The \textbf{permuted sequential MNST} has been frequently used to test the ability of RNNs to learn long-term dependencies.
 Compared to the original task, image data is vectorized and presented sequentially in a fixed, random order.
 The sequential presentation, and the non-local structure induced by the permutation, require that the network use long-range dependencies to successfully classify the input.
The maximal number of epochs allotted for training was $125$ .

\subsection[Synthetic]{Synthetic datasets}
All of the synthetic datasets described below were generated to have $10,000$ training samples, $1,000$ validation samples, and $1,000$ test samples.
To ensure a fair comparison, the dimensionality of the RNN state was set to $250$ for all of the synthetic datasets.
The dimensionality is comparable to those used in~\citet{Arjovsky2016}.
The maximum number of epochs allotted for training was $400$.
Cross-entropy was used as the loss function, and we report the average loss over the test set.

\subsubsection{Copy memory}
The copy-memory task was first introduced in~\citep{Hochreiter1997}.
It requires a model to retain an input sequence and then reproduce it as the output following a $k$-step delay.
The input consists of a sequence of $10$ tokens drawn at random from an alphabet of size $8$, followed by $k$ repetitions of a `blank' and a single `start' token, and $9$ `blank' tokens.
The target output is a sequence of $k+10$ `blank' tokens followed by the original $10$ element-long sequence presented at the beginning of the input.

\subsubsection{Temporal Ordering}
The input consists of a sequence of discrete symbols, which emit a symbol from the set of {A, B} at the beginning and throughout the sequence.
The objective of the task is to classify the order of these symbols (i.e. AA, AB, BA, or BB) following a $k$-step delay, where $k\sim\mathcal{U}\{50,60\}$.

\subsubsection{Random-delay copy memory}
The random-delay copy-memory task is a variant of the copy-memory task where the delay was drawn randomly from a uniform distribution between $100$ and $120$ steps.

\subsection{Hyperparameter selection}
The process of hyperparameter tuning was performed using the Ray Tune library with Skopt as the search algorithm.
The following hyperparameters were optimized:
\begin{itemize}
  \item \textbf{ADAM learning rate} optimized on a log-scale with ranges $10^{-5}$ to $1$.
  \item \textbf{Gradient clipping} optimized on a log-scale with ranges $10^{-2}$ to $10^2$.
  \item \textbf{Dropout} optimized on a log-scale with ranges $10^{-1.5}$ to $0.9$.
  \item \textbf{Learning rate decay factor} optimized on a log-scale with ranges $10^{-1}$ to $1$.
  \item \textbf{Early stopping}, on a range from $5$ to $80$.
\end{itemize}
For architectures with additional hyperparameters, such as \textbf{coRNN} and \textbf{LipschitzRNN}, hyperparameter tuning was performed in ranges reported in their respective, original publications~\citep{Rusch2021,Erichson2021}.
The tuning of all hyperparameters was performed using $200$ trials, except the permuted sequential MNST task, where $100$ trials were used.
All tuning was performed on a validation set that was subsampled from the training set.

\begin{table*}\label{tab:allresults}
\resizebox{\textwidth}{!}
{
\centering
\begin{tabular}{lccccccccc}
\toprule
 {} & \multirow{3}{*}{\small Antisym } & \multirow{3}{*}{ \small coRNN} & \multicolumn{2}{c}{GRU} & \multirow{3}{*}{\small LSTM} & \multirow{3}{*}{Lip.} & \multicolumn{3}{c}{\small $\tanh$ RNN} \\
 \cline{4-5}\cline{8-10}
 {} & {} & {} & \multirow{2}{*}{\small def.} & \multirow{2}{*}{\lcglyph} & {} & {} & \multirow{2}{*}{def.} & \multirow{2}{*}{\lcglyph} & \multirow{2}{*}{\small $W\trp W = I$} \\
{} & {} & {} & {} & {} & {} & {} & {} & {} & {} \\
\midrule
\rowcolor{gray!20}
\small Temporal Ordering $\downarrow$ & -- & -- & $\mathbf{0}$ & $\mathbf{0}$ & $1.11$ & -- & $1.39$ & $1.22 \times 10^{{-5}}$ & $0.31$ \\
\small Copy memory $k = 100$ $\downarrow$& $0.30$ & $0.11$ & $0.29$ & $\mathbf{1.33 \times 10^{{-4}}}$ & $1.18$ & $0.03$ & $1.86$ & $1.58 \times 10^{{-4}}$ & $1.35$ \\
\rowcolor{gray!20}
\small Copy memory $k = 2000$ $\downarrow$& $1.91$ & $1.87$ & $0.22$ & $0.35$ & $1.61$ & $1.91$ & $1.99$ & $\mathbf{0.05}$ & $1.98$ \\
\small Random-delay copy memory $\downarrow$& $0.74$ & $0.40$ & $0.72$ & $\mathbf{1.49 \times 10^{{-7}}}$ & $1.57$ & $0.02$ & $1.85$ & $0.33$ & $1.55$ \\
\rowcolor{gray!20}
\small Permuted Seq. MNIST $\uparrow$& $95.9$ & $94.5$ & $94.8$ & $\mathbf{97.4}$ & $91.6$ & $94.9$ & $88.8$ & $97.0$ & $94.7$ \\
\bottomrule
\end{tabular}
}
    \caption{
      Test set performance of PTA initializations and other recurrent models on a gamut of benchmarks.
      The reported values are \textbf{averages over 5 model realizations}.
    }
\end{table*}

\end{document}